\input amstex
\magnification 1200
\TagsOnRight
\def\qed{\ifhmode\unskip\nobreak\fi\ifmmode\ifinner\else
 \hskip5pt\fi\fi\hbox{\hskip5pt\vrule width4pt
 height6pt depth1.5pt\hskip1pt}}
\NoBlackBoxes
\baselineskip 20 pt
\parskip 5 pt
\def\stretch {\noalign{\medskip}}
\define \bCp {\bold C^+}
\define \bCm {\bold C^-}
\define \bCpb {\overline{\bold C^+}}
\define \bCmb {\overline{\bold C^-}}
\define \ds {\displaystyle}
\define \bR {\bold R}
\define \bm {\bmatrix}
\define \endbm {\endbmatrix}

\centerline {\bf THE UNIQUENESS IN THE INVERSE PROBLEM}
\centerline {\bf FOR TRANSMISSION EIGENVALUES} \centerline {\bf
FOR THE SPHERICALLY-SYMMETRIC} \centerline {\bf VARIABLE-SPEED
WAVE EQUATION}

\vskip 7 pt
\centerline {Tuncay Aktosun}
\vskip -8 pt
\centerline {Department of Mathematics}
\vskip -8 pt
\centerline {University of Texas at Arlington}
\vskip -8 pt
\centerline {Arlington, TX 76019-0408, USA}
\vskip -8 pt
\centerline {aktosun\@uta.edu}

\centerline {Drossos Gintides and Vassilis G. Papanicolaou}
\vskip -8 pt
\centerline {Department of Mathematics}
\vskip -8 pt
\centerline {National Technical University of Athens}
\vskip -8 pt
\centerline {Zografou Campus}
\vskip -8 pt
\centerline {157 80, Athens, Greece}
\vskip -8 pt
\centerline {dgindi\@math.ntua.gr and papanico\@math.ntua.gr}

\noindent {\bf Abstract}: The recovery of a spherically-symmetric wave speed $v$ is considered in a bounded spherical
region of radius $b$
 from the set of the corresponding transmission
eigenvalues for which the corresponding eigenfunctions are
also spherically symmetric. If the integral of $1/v$ on the interval
$[0,b]$ is less than $b,$ assuming that there exists at least one
$v$ corresponding to the data, it is shown that $v$ is uniquely
determined by the data consisting of such transmission
eigenvalues and their ``multiplicities," where the
``multiplicity" is defined as the multiplicity of the transmission
eigenvalue as a zero of a key quantity. When that integral is
equal to $b,$ the unique recovery is obtained when the data
contains one additional piece of information. Some similar
results are presented for the unique determination of the
potential from the transmission eigenvalues with
``multiplicities" for a related Schr\"odinger equation.

\vskip -8 pt
\par\noindent {\bf Short title:} Inverse problem for transmission eigenvalues
\vskip -8 pt
\par\noindent {\bf Keywords:}
transmission eigenvalues, inverse spectral problem,
variable-speed wave equation, Schr\"odinger equation

\newpage

\noindent {\bf 1. INTRODUCTION}
\vskip 3 pt

The interior transmission problem is a nonselfadjoint
boundary-value problem for a pair of fields $\Psi$ and $\Psi_0$
in a bounded and simply connected domain $\Omega$ of $\bR^n$
with the sufficiently smooth boundary $\partial \Omega.$ It can be formulated as
$$
\cases
\Delta \Psi +\lambda\,\rho(\bold x)\,\Psi=0,\qquad \bold x\in\Omega,\\
\stretch
\Delta\Psi_0+\lambda\Psi_0=0,\qquad\bold x\in\Omega,\\
\stretch \Psi=\Psi_0,\quad \displaystyle\frac{\partial
\Psi}{\partial\bold n}=\frac{\partial\Psi_0}{\partial\bold
n}, \qquad \bold x\in \partial \Omega, \endcases \tag 1.1$$
where $\Delta $ denotes the Laplacian, $\lambda$ is the
spectral parameter, $\bold n$ represents the outward unit
normal to the boundary $\partial \Omega,$ and the positive
quantity $\rho(\bold x)$ corresponds to the square of the
refractive index of the medium at location $\bold x$ in the
electromagnetic case or the reciprocal of the square of the
sound speed $v(\bold x)$ in the acoustic case, i.e. $v(\bold
x):=1/\sqrt{\rho(\bold x)}.$ In the acoustic case,
$\sqrt{\rho(\bold x)}$ is usually called the slowness. Without
loss of generality we can assume that in the region exterior to
$\Omega,$ the speed of the electromagnetic wave is $1$ or the
sound speed is $1$ in the acoustic case.

This interior transmission problem arises in the
inverse scattering theory in inhomogeneous media,
where the goal is to determine the function $\rho$
in $\Omega$ from an appropriate set of $\lambda$-values related to (1.1).
The values of $\lambda $ for which (1.1) has a pair
of nontrivial solutions $\Psi$ and
$\Psi_0$ are called {\it transmission eigenvalues.} It is already
known that those transmission eigenvalues can be
determined from some far-field measurements (see e.g.
[5,7,13-15,31]).

Since there is not a standard theory to analyze nonselfadjoint
eigenvalue problems, the existence of transmission eigenvalues
for (1.1) was an open problem until recently. Using some techniques
related to the Fredholm theory of integral equations, it has
been shown [14] that the transmission eigenvalues for (1.1)
form a discrete set with infinity as the only possible
accumulation point. In general we expect transmission
eigenvalues to be complex numbers although some of them may be real
and some, in fact, may be positive.

Under the assumption that
$\rho(\bold x)\ge 1$ for all $\bold x\in \Omega$ (or the assumption
that $\rho(\bold x)\le 1$ for all $\bold x\in \Omega$), it has been
shown [11] that the corresponding positive transmission
eigenvalues for (1.1) form a countably infinite set. Similar results for the existence of a countably
infinite set of transmission eigenvalues have been obtained
[9,10,12,18,28] for related problems involving Helmholtz and
Maxwell's equations, where the bounded region $\Omega$ is
allowed to contain cavities, where $\rho(x)\equiv 1$
in each cavity.

A fundamental problem related to (1.1) is the relationship
between $\rho$ in $\Omega$ and the corresponding
transmission eigenvalues. In [4,8] it has been observed that the
transmission eigenvalues
carry some information about $\rho$ in $\Omega.$
The case $n=3$ is
naturally the most relevant in applications.
A key question
is whether we can uniquely determine $\rho$ in $\Omega$ if all the
transmission eigenvalues are known. Another important question is whether
the unique recovery is possible if we know only a certain subset
of the transmission eigenvalues.

In the case where $\Omega$ is the ball of radius $b$ and
$\rho(\bold x)$ is radially symmetric, it has recently been
shown [9] that the set of all transmission eigenvalues uniquely determine
$\rho$ in $\Omega.$ In the radially symmetric case, let us use
$\rho(x)$ instead of $\rho(\bold x)$ with $x:=|\bold x|.$ In
this case it is natural to ask whether $\rho(x)$ can be
determined from a subset of transmission eigenvalues, such as
those transmission eigenvalues for which the corresponding
eigenfunctions are also spherically symmetric. We will refer to
such eigenvalues as {\it special transmission eigenvalues.}
Another variant of the transmission eigenvalue problem in the
spherically-symmetric case has been studied in [24-26], where
some uniqueness results were established when only the positive
special transmission eigenvalues are used in the determination.

In the case $n=3,$
where $\Omega$ is the ball of radius $b > 0$ and $\rho$ is radially
symmetric, the boundary-value problem (1.1)
becomes equivalent to a nonstandard Sturm-Liouville-type
eigenvalue problem,
which is formulated in the following
proposition. Here, ``nonstandard"
refers to the fact that the spectral parameter appears in
the boundary condition at the right endpoint.
Our assumptions
on $\rho$ are that $\rho(x)$ is positive and continuously differentiable
and that
$\rho''$ is square integrable, i.e.
$$
\rho (x) > 0,\quad x\in(0,b);\quad
\rho \in C^{1}(0,b);\quad \rho''\in L^2(0,b).
\tag 1.2$$
where a prime is used to denote the $x$-derivative.

\noindent {\bf Proposition 1.1} {\it Consider the
special case of (1.1) with
$\Omega$ being the three-dimensional ball of
radius $b$ centered at the origin,
where only
spherically-symmetric wave functions are allowed
and it is assumed that such wave functions
are continuous in the closure of $\Omega.$ Then, the
corresponding special transmission eigenvalues of
(1.1) coincide with the eigenvalues of the
nonstandard boundary-value problem}
$$\cases\Phi''+\lambda\, \rho (x)\,\Phi=0,\qquad 0<x<b,\\
\stretch
\Phi(0) = 0, \quad
\displaystyle\frac{\sin (\sqrt{\lambda }\,b)}{\sqrt{\lambda }}\,
\Phi'(b)-\cos (\sqrt{\lambda }\,b)\,\Phi(b)=0.\endcases  \tag 1.3$$

\noindent PROOF: The Laplacian in $\bR^3$ in polar coordinates
$(x,\theta,\varphi)$ is given by
$$\Delta:=\ds\frac{1}{x^2}\ds\frac{\partial}{\partial x}
x^2\ds\frac{\partial}{\partial x}+\ds\frac{1}
{x^2\sin\varphi}\ds\frac{\partial}{\partial \varphi}\sin\varphi
\ds\frac{\partial}{\partial \varphi}+
\ds\frac{1}
{x^2\sin^2\varphi}
\ds\frac{\partial^2}{\partial \theta^2},\tag 1.4$$
where we recall that $x:=|\bold x|.$
If the wave functions
$\Phi$ and $\Phi_0$ are spherically symmetric, i.e. if
they do not depend on $\theta$ and $\varphi,$ then
with the help of (1.4) we transform (1.1) into
$$\cases
\Psi''+\ds\frac{2\Psi'}{x}+\lambda\,\rho(x)\,\Psi=0,\qquad 0<x<b,\\
\stretch
\Psi_0''+\ds\frac{2\Psi_0'}{x}+\lambda\,\Psi_0=0,\qquad 0<x<b,\\
\stretch
\Psi(b)=\Psi_0(b), \quad \Psi'(b)=\Psi'_0(b),\endcases\tag 1.5$$
where $\Psi(0)$ and $\Psi_0(0)$ must be finite
because of the continuity of
$\Psi$ and $\Psi_0$ in $\Omega.$
Letting $\Phi:=x\Psi$ and $\Phi_0:=x\Psi_0,$ from (1.5) we get
$$\cases
\Phi''+\lambda\,\rho(x)\,\Phi=0,\qquad 0<x<b,\\
\stretch
\Phi_0''+\lambda\,\Phi_0=0,\qquad 0<x<b,\\
\stretch
\Phi(0)=\Phi_0(0)=0,
\quad \Phi(b)=\Phi_0(b), \quad \Phi'(b)=\Phi'_0(b).\endcases\tag 1.6
$$
 From the second line in (1.6) we see that the solution $\Phi_0(x)$
satisfying
$\Phi_0(0)=0$ must be a constant multiple of $\sin(\sqrt{\lambda}\,x)/\sqrt{\lambda}.$
Thus, we see that (1.6) is equivalent to (1.3). \qed

The eigenvalues of (1.3), namely the
$\lambda$-values for which (1.3) has a nontrivial solution $\Phi(x)$,
are the special transmission eigenvalues mentioned earlier.
In other words, the corresponding eigenfunctions are spherically symmetric
and hence functions of $x$ only.
Note that
such eigenfunctions of (1.3)
can only be determined up to a multiplicative
constant, and it is clear from
(1.3) that there exists only one linearly
independent eigenfunction for each eigenvalue
of (1.3). Nevertheless, for each eigenvalue $\lambda_j$
of (1.3) we will associate a ``multiplicity"
in a special sense, namely the multiplicity of
$\lambda_j$ as a zero of the quantity
$D(\lambda)$ defined in (2.10).
We will elaborate on the meaning of ``multiplicity" in Section~2.

We define the relevant quantity $a$ as
$$
a:=\int_{0}^{b}\sqrt{\rho (x)}\,dx,\tag 1.7$$
which has the physical interpretation as the travel
time for the wave to move from $x=0$ to $x=b.$
Our main result in this paper is the
proof that the knowledge of eigenvalues of (1.3) with their
``multiplicities" uniquely determine $\rho (x)$ for $0<x<b$
provided $a<b.$ If $a=b,$ we prove the unique determination of
$\rho$ provided we know one additional parameter, namely the
value of the constant $\gamma$ appearing in (2.13).
Let us clarify
that we do not study the existence aspect of the inverse
problem but we only analyze the uniqueness aspect. In other
words, corresponding to our data we assume that there exists at
least one function $\rho$ satisfying (1.2), and we prove that
if $\rho_1$ and $\rho_2$ are two such functions then we must
have $\rho_1\equiv\rho_2.$

When $a=b,$
it is an open question if knowledge of $\gamma$ is necessary or
whether $\gamma$ can be determined from the knowledge of
eigenvalues of (1.3) including their ``multiplicities." In the
discrete version of (1.3), assuming the existence aspect of
the inverse problem is solved, it is already known [29] that
generically, except for one exceptional case, $\rho$ is uniquely
determined from the knowledge of the special transmission
eigenvalues and their ``multiplicities" and hence $\gamma$ is
in general uniquely determined without needing to know any
additional parameter.

If $\rho (x)$
satisfies (1.2), it is known [24,26] that
$$\lambda _{n_{j}}=\ds\frac{n_{j}^{2}\pi^{2}}{(a-b)^{2}}+O(1),
\qquad n_j \to +\infty,$$ where $\lambda _{n_{j}}$ for
$j\in\bold N$ are the real eigenvalues of (1.3) indexed in an
increasing order, with $\bold N$ denoting the set of
positive integers. Hence, the quantity $a$ can be determined if
the real eigenvalues of (1.3) are known. In other words, if
$\rho_1$ and $\rho_2$ satisfy (1.2) and they correspond to the
same set of special transmission eigenvalues, then we must have
$a_1=a_2,$ where
$$a_1:=\int_{0}^{b}\sqrt{\rho_1(x)}\,dx,\quad
a_2:=\int_{0}^{b}\sqrt{\rho_2(x)}\,dx.\tag 1.8$$

Let us elaborate on the eigenvalues of (1.3). As we illustrate
with some examples in Section~2, besides real eigenvalues,
(1.3) has in general nonreal eigenvalues and in fact the number
of nonreal eigenvalues may be infinite. Because $\rho(x)$ is
real valued, from (1.3) it is seen that if $\lambda$ is an
eigenvalue then $\lambda^*$ is also an eigenvalue of (1.3),
where we use an asterisk to denote complex conjugation. In our
present work, for the unique recovery of $\rho$ we assume
the knowledge of all the eigenvalues (both real and complex
nonreal) including their ``multiplicities." In the previously
established uniqueness results [24-26] regarding (1.3) it has
been assumed that either $a \leq b/3$ or that  some partial
information on $\rho$ is available. On the other hand, in those results
[24-26] it is
assumed that only the positive eigenvalues are known and no
``multiplicities" are used in the data.

Our paper is organized as follows. In Section~2 we present
some preliminary results that are needed to prove
the uniqueness theorems of Sections~3 and 4.
In Section~3 we consider
the uniqueness in the recovery of
$\rho$ from the knowledge of
special transmission eigenvalues of (1.1)
with ``multiplicities." When $a<b,$ where
$a$ is the quantity in (1.7), we establish the uniqueness.
When $a=b,$ we show that the combined knowledge of
special transmission eigenvalues of (1.1)
with ``multiplicities" and the constant $\gamma$ appearing
in (2.13) assures the uniqueness. We also elaborate on the
case $a>b$ and indicate why the technique we use
does not apply in that case to prove the uniqueness.
In Section~4 we consider the uniqueness in the
recovery of the potential $V$ of the Schr\"odinger
equation from the data consisting of
special transmission eigenvalues of (4.1)
with ``multiplicities." We prove the unique recovery
if our data contains one additional parameter, namely the constant
$\tilde\gamma$ appearing in (4.5).

\vskip 10 pt
\noindent {\bf 2. PRELIMINARIES}
\vskip 3 pt

Let us recall [1] that an entire function of order $1/2$
grows no faster than $O(e^{c\,|\lambda|^{(1/2)+\epsilon}})$ as
$\lambda\to\infty$ in the complex plane $\bold C$ for any given
positive $\epsilon,$ where $c$ is some positive constant. The
sums and products of such functions are entire of order not
exceeding $1/2.$

We first consider a problem closely related to (1.3), namely
$$\cases\phi''+\lambda\, \rho (x)\,\phi=0,\qquad 0<x<b,\\
\stretch \phi(0) = 0, \quad \phi'(0)=1.\endcases  \tag 2.1$$ It
is known [30] that, for every $\lambda$ in the
complex plane $\bold C,$ (2.1) has a
unique solution $\phi(x),$ which we also write as
$\phi(x;\lambda)$ to emphasize its dependence on $\lambda.$
Since $\rho(x)$ is real valued, the solution to (2.1) satisfies
$$\phi(x;\lambda^*)=\phi(x;\lambda)^*,
\qquad \lambda\in\bold C.\tag 2.2$$

\noindent {\bf Proposition 2.1} {\it Assume that
$\rho$ satisfies (1.2). Then, (2.1) is uniquely solvable,
and for each fixed $x\in(0,b]$
the quantities $\phi(x;\cdot)$ and
$\phi'(x;\cdot)$ are entire in $\lambda$ of order $1/2.$
Furthermore,
$\phi(x;\lambda)$ and $\phi'(x;\lambda)$
cannot simultaneously vanish at the same $x$-value.}

\noindent PROOF: We refer the reader to [30]
for the proof that $\phi(x;\cdot)$ and
$\phi'(x;\cdot)$ are entire in $\lambda$ of order $1/2.$
If $\phi(x_0;\lambda)=\phi'(x_0;\lambda)=0$ for some
$x_0$ value in $[0,b],$ then the unique solution
to the corresponding
initial-value problem would have to be the
zero solution, which is incompatible with
$\phi'(0)=1$ in (2.1). \qed

When $\rho$ satisfies (1.2),
it is known (see e.g. [30]) that the variable-speed
wave equation in (2.1) can be transformed
into a Schr\"odinger equation via a Liouville transformation.
In other words, by using the change of variables
$$
y=y(x):=\int_{0}^{x}\sqrt{\rho (s)}\,ds ,\quad \varphi(y)
=\varphi(y(x) ):=\rho
(x)^{1/4}\,\phi(x),\tag 2.3$$
we can transform (2.1) into the equivalent Sturm-Liouville problem
for the Schr\"odinger equation that is given by
$$\cases
-\varphi''(y)+q(y)\,\varphi(y)=\lambda\,
\varphi(y),
\qquad 0<y<a,\\
\stretch
\varphi(0)=0,\quad \varphi'(0)=\ds\frac{1}{\rho (0)^{1/4}},
\endcases\tag 2.4$$
where $a$ is the quantity defined in (1.7) and
$$
q(y)=
q(y(x)):=\ds\frac{1}{4}\frac{\rho ^{\prime \prime }(x)}{\rho (x)^{2}}-\frac{5}{16
}\frac{\rho ^{\prime }(x)^{2}}{\rho (x)^{3}}.$$

Let us use $\text{Im}[\sqrt{\lambda }]$ to denote the
imaginary part of $\sqrt{\lambda },$ where the argument of the
square-root function is chosen so that $\arg(\sqrt{\lambda})\in(-\pi/2,\pi/2].$
The proof of the following
proposition can be obtained [30] with the help
of the Liouville transformation (2.3) and some estimates
for the corresponding Schr\"odinger equation in (2.4), and hence
it will not be given here.

\noindent {\bf Proposition 2.2} {\it Assume that $\rho$
satisfies (1.2). Then there exists a positive constant $A$ such
that, for all $x\in [0,b]$ and $\lambda \in \bold
C,$ the solution $\phi(x;\lambda)$ to (2.1) and
its $x$-derivative, respectively, satisfy}
$$
\left|\phi(x;\lambda )-\frac{1}{\left[ \rho (0)\,\rho (x)\right] ^{1/4}\sqrt{
\lambda }}\,\sin \left( \sqrt{\lambda }\,y(x)
\right) \right| \leq \frac{A}{\left|\sqrt{\lambda }\right|}\exp \left( \left|\text{Im}[\sqrt{
\lambda }]\right|\,y(x) \right),\tag 2.5$$
$$\left|\phi'(x;\lambda )-\left[ \frac{\rho (x)}{\rho (0)}\right]
^{1/4}\cos \left( \sqrt{\lambda }\,y(x) \right)
\right| \leq A\exp \left( \left|\text{Im} [\sqrt{\lambda }]\right|
\,y(x) \right) ,\tag 2.6$$
{\it where $y(x)$ is the quantity given in (2.3).}

For a positive $\varepsilon$ let
$\bold C_{\varepsilon}$ denote the sector in the complex plane
defined as
$$
\bold C_{\varepsilon}:=\{\lambda\in \bold C:\
\varepsilon\le\arg(\lambda)\le 2\pi -\varepsilon\}.\tag 2.7$$
The proof of the following result is already known [27].

\noindent {\bf Proposition 2.3} {\it
Assume that $\rho$ satisfies (1.2).
Then, for each fixed $x\in [0,b],$ as $\lambda\to\infty$
in $\bold C_{\varepsilon},$
the unique solution $\phi(x;\lambda)$ to (2.1) satisfies}
$$
\phi(x;\lambda )=\frac{1}{\left[ \rho (0)\,\rho (x)\right] ^{1/4}\sqrt{\lambda }}
\left[ \sin \left( \sqrt{\lambda }\,y(x) \right) \right]\left[
1+O\left( \frac{1}{\sqrt{\lambda }}\right) \right],\tag 2.8$$
$$
\phi'(x;\lambda )=\left[ \frac{\rho (x)}{\rho (0)}\right] ^{1/4} \left[ \cos
\left( \sqrt{\lambda }\,y(x) \right) \right] \left[
1+O\left( \frac{1}{\sqrt{\lambda }}\right) \right],\tag 2.9$$
{\it where $y(x)$ is the quantity given in (2.3).}

Let us now clarify the relationship between (1.3) and (2.1).
In general, for a given $\lambda\in\bold C,$ (1.3) may not
have a nontrivial solution.
Suppose
that $\lambda_j$ is an eigenvalue of (1.3). Then a solution
$\Phi(x;\lambda_j)$ to (1.3) can only be determined
up to a multiplicative constant, and in fact any such solution must be
a constant multiple of the unique solution $\phi(x;\lambda_j)$ to (2.1)
due to the fact that $\Phi(0)=0$ in (1.3)
and $\phi(0)=0$ in (2.1).

We now introduce the key function $D(\lambda)$ as
$$
D(\lambda):=\frac{\sin(\sqrt{\lambda }\,b)}{\sqrt{\lambda}}
\,\phi'(b;\lambda )-\cos (\sqrt{\lambda}\,b)\,\phi(b;\lambda),\tag 2.10$$
where we recall that $\phi(x;\lambda)$ is the unique solution to (2.1).
Let us remark that, if $\rho(x)\equiv 1$ in (2.1), then
$\phi(x;\lambda)=\sin(\sqrt{\lambda}\,x)/\sqrt{\lambda}$ and hence
$D(\lambda)\equiv 0.$

\noindent {\bf Theorem 2.4} {\it Assume that $\rho$ satisfies
(1.2). Then, the quantity $D(\lambda)$ defined in (2.10) is
entire in $\lambda$ of order not exceeding $1/2.$ Each zero of
$D(\lambda)$ in the complex plane $\bold C$ corresponds to an
eigenvalue of (1.3) and vice versa. The value $\lambda=0$ is
always a zero of $D(\lambda)$ of some multiplicity $d$ with
$d\ge 1,$ and hence}
$$D (0)=0.\tag 2.11$$
{\it Furthermore,}
$$D(\lambda^*)=D(\lambda)^*,\qquad
\lambda\in\bold C,\tag 2.12$$
{\it and there exists a real constant $\gamma$ such that}
$$D (\lambda )=\gamma\, \Xi(\lambda ),\tag 2.13$$
{\it where the auxiliary quantity $\Xi(\lambda )$ is
uniquely determined from the zeros (including multiplicities) of $D(\lambda)$
and has the
representation}
$$
\Xi(\lambda )=\lambda ^{d}\prod_{n=1}^{\infty }\left( 1-\frac{
\lambda }{\lambda _{n}}\right),\tag 2.14$$
{\it with $\lambda_n$ for $n\in\bold N$
being the nonzero
zeros of
$D(\lambda),$ some of which may be repeated.}

\noindent PROOF: From their representations in terms of exponential
functions, we know that $\sin(\sqrt{\lambda}\,b)/\sqrt{\lambda}$
and $\cos(\sqrt{\lambda}\,b)$ are entire in $\lambda$ of order $1/2.$
 From Proposition~2.1 we know that $\phi(b;\lambda)$ and
$\phi'(b;\lambda)$ are entire of order $1/2,$ and hence the
right side of (2.10) is entire of order not exceeding $1/2.$ If
$\lambda_j$ is an eigenvalue of (1.3) with an eigenfunction
$\Phi(x;\lambda_j),$ we already know that $\Phi(x;\lambda_j)$
is a constant multiple of the solution $\phi(x;\lambda_j)$ to
(2.1), and hence from (2.10) we see that $D(\lambda_j)=0.$
Conversely, if $D(\lambda_j)=0$ for some
$\lambda_j,$ then a comparison of (1.3) and (2.1) shows that
the unique solution $\phi(x;\lambda_j)$ to (2.1) satisfies
(1.3) and hence $\lambda_j$ is an eigenvalue for (1.3) with
eigenfunction $\phi(x;\lambda_j).$ In particular, we note that
when $\lambda=0$ the unique
solution to (2.1) is given by
$$\phi(x;0)=x,\tag 2.15$$
which indicates that
$$\phi(0;0)=0,\quad \phi'(0;0)=1,
\quad \phi(b;0)=b,\quad \phi'(b;0)=1,\tag 2.16$$
and hence $\phi(x;0)$ indeed satisfies
(1.3) when $\lambda=0.$ Thus, $\lambda=0$ is always a zero
of $D(\lambda)$ with some multiplicity $d,$
which is at least one. We obtain (2.12) from (2.2) and (2.10).
Since $D(\lambda)$ is entire of
order not exceeding $1/2,$ by the Hadamard
factorization theorem, we must have
the representation in (2.13),
where $\gamma$ is a complex constant and
$\Xi(\lambda)$ as in
(2.14). In fact
$\gamma$ turns out to be real as a result of (2.12). \qed

As we have seen in Proposition~1.1 and Theorem~2.4, the special
transmission eigenvalues of (1.1),
the eigenvalues of (1.3), and the zeros of
$D(\lambda)$ defined in (2.10) all coincide.
On the other hand, each zero of
$D(\lambda)$ may have a multiplicity greater than one even though
there exists only one linearly independent
eigenfunction for the corresponding eigenvalue of (1.3).
We refer to the multiplicity of a zero $\lambda_j$
of $D(\lambda)$ also as the ``multiplicity" of
the special transmission eigenvalue $\lambda_j.$
Next, we elaborate on the ``multiplicities"
with an illustrative example.

\noindent {\bf Example 2.5} When $\rho(x)$ is constant on
$(0,b),$ by using $\rho$ to denote that constant value, the
unique solution to (1.3) is obtained as
$$\phi(x;\lambda)=\ds\frac{1}{\sqrt{\lambda\rho}}
\,\sin(\sqrt{\lambda\rho}\,x),\qquad
0<x<b,$$
and hence the corresponding quantity in (2.10) is given by
$$D(\lambda)=\ds\frac{1}{\sqrt{\lambda}}
\,\sin(\sqrt{\lambda}\,b)\,\cos(\sqrt{\lambda\rho}\,b)-
\ds\frac{1}{\sqrt{\lambda\rho}}\,\cos(\sqrt{\lambda}\,b)\,\sin(\sqrt{\lambda\rho}\,b).\tag
2.17$$ When $\rho(x)\equiv 1/4,$ from (2.17) we get
$$D(\lambda)=\ds\frac{2}{\sqrt{\lambda}}\,\sin^3\left(\ds\frac{\sqrt{\lambda}\,b}{2}\right),
$$ and hence $D(\lambda)$ has a simple zero at
$\lambda=0$ and an infinite set of real zeros at the
$\lambda$-values $4j^2\pi^2/b^2$ for $j\in\bold N,$ each having
a multiplicity of three. On the other hand, when $\rho(x)\equiv
4/9,$ from (2.17) we get
$$D(\lambda)=\ds\frac{1}{\sqrt{\lambda}}\,\sin^3\left(\ds\frac{\sqrt{\lambda}\,b}{3}\right)
\left[3+2\,\cos\left(\ds\frac{2\sqrt{\lambda}\,b}{3}\right)
\right],$$
and hence $D(\lambda)$ has a simple zero at $\lambda=0,$ an infinite set of real zeros of
multiplicity three at the $\lambda$-values
$9j^2\pi^2/b^2$ for $j\in\bold N,$ and an infinite set of simple complex zeros
at the $\lambda$-values that are given by
$$\ds\frac{9(2j-1)^2\pi^2}{4b^2}-
\ds\frac{9}{4b^2}\left[\log\left(\ds\frac{3+\sqrt{5}}{2}\right)\right]^2
\pm i\,\ds\frac{9(2j-1)\pi}{2b^2}\left[\log\left(\ds\frac{3+\sqrt{5}}{2}
\right)\right],
\qquad j\in\bold N.$$

Let us remark that the knowledge of $\Xi(\lambda)$ given in
(2.15) is equivalent to the knowledge of the eigenvalues of
(1.3) with their ``multiplicities." Furthermore, the knowledge
of $\Xi(\lambda)$ is equivalent to the knowledge of its zeros
including their multiplicities. Hence, in proving our
uniqueness results, as our data we can equivalently use
$\Xi(\lambda),$ the zeros of $\Xi(\lambda)$ with their
multiplicities, the eigenvalues of (1.3) with their
``multiplicities," or the special transmission eigenvalues
of (1.1)
with their ``multiplicities."

Since $D(\lambda)$ given in (2.10) is entire, we can obtain its
Maclaurin expansion with the help of the Maclaurin expansion of
the unique solution $\phi(x;\lambda)$ to (2.1), which we
write as
$$\phi(x;\lambda)=\phi_0(x)+\lambda\,\phi_1(x)+\lambda^2\phi_2(x)+O(\lambda^2),
\qquad \lambda\to 0 \text{ in }
\bold C,\tag 2.18$$
where we have defined
$$\phi_0(x):=x,\quad \phi_1(x):=M_2(x)-x\,M_1(x),\tag 2.19$$
$$\phi_2(x):=\ds\frac{1}{2}\int_0^x dz\,[M_1(z)]^2-x\int_0^x dz\,z\,M_2(z)+\int_0^x
dz\,z\,\rho(z)\,M_2(z),\tag 2.20$$
with
$$M_1(x):=\int_0^x dz\,z\,\rho(z),\quad
M_2(x):=\int_0^x dz\,z^2\,\rho(z).$$ Using
(2.18)-(2.20) and their $x$-derivatives in (2.10) we obtain
$$D(\lambda)=D_0+\lambda\,D_1+\lambda^2 D_2+
O(\lambda^3),
\qquad \lambda\to 0 \text{ in }
\bold C,\tag 2.21$$
where
$$D_0:=0,\quad D_1:=\ds\frac{b^3}{3}-M_2(b),$$
$$D_2:=-\ds\frac{b^5}{30}+b\,[M_1(b)]^2-M_1(b)\,M_2(b)-\ds\frac{b^3}{3}\,M_1(b)+
\ds\frac{b^2}{2}\,M_2(b)-\int_0^b dz\,[M_1(z)]^2.$$
If $d=1$ in (2.14), with the help of
(2.10) and (2.21) we see that
$$D_1\ne 0,\quad \gamma=D_1,
\quad -\gamma\sum_{j=1}^\infty \ds\frac{1}{\lambda_j}=D_2,$$
where $\gamma$ is the parameter appearing in (2.13) and
$\lambda_j$ for $j\in\bold N$ are the nonzero zeros of $D(\lambda),$ some
of which may be repeated. On the other hand, if $d=2$ in (2.14), then we must have
$$D_1=0,\quad
D_2\ne 0,\quad \gamma=D_2.$$

The results in the following propositions will be used in the proof
of the unique determination of $\rho.$

\noindent {\bf Proposition 2.6} {\it Suppose that $h$ is an entire
function of $\lambda,$ and let}
$$f(\lambda):=\ds\frac{\sin(\sqrt{\lambda}\,c)}{\sqrt{\lambda}}\,h(\lambda),\tag 2.22$$
{\it for some positive constant $c,$
and assume that $f(\lambda)$ is entire of order not exceeding
1/2.
Then, the order of $h(\lambda)$ cannot exceed 1/2.}

\noindent PROOF: Note that (2.22) and the fact that the order of
$f$ does not exceed $1/2$ imply that for any positive $\epsilon$ there exists
a positive constant $A$ such that
$$|h(\lambda)|\le
\ds\frac{|\sqrt{\lambda}|}{|\sin(\sqrt{\lambda}\,c)|}
\,|f(\lambda)|\le
\ds\frac{A}{|\sin(\sqrt{\lambda}\,c)|}\,
\exp\left(c\,|\lambda|^{(1/2)+\epsilon}\right).\tag 2.23$$
In the neighborhood of the zeros of $\sin(\sqrt{\lambda}\,c)/\sqrt{\lambda},$
which occur when
$\lambda=n^2\pi^2/c^2$ for $n\in\bold N,$ the bound
in (2.23) is too large to assure that
the order of $h(\lambda)$ cannot exceed $1/2.$
Thus, we need to analyze the behavior of
$h(\lambda)$ near those zeros.
Let us enclose each
such zero within the disk $U_n$ of radius one, where we have
defined
$$U_n:=\left\{\lambda\in\bold C:\ \left|\lambda-\ds\frac{n^2\pi^2}{c^2}\right|<1\right\},
\qquad n\in\bold N.$$ The boundary $\partial U_n$ can
be parameterized by using $\vartheta$ so that if
$\lambda\in \partial U_n$ then
$$\lambda=\ds\frac{n^2\pi^2}{c^2}+e^{i\vartheta},\qquad \vartheta\in(-\pi,\pi],$$
or equivalently
$$\lambda c^2=n^2\pi^2\left(1+\ds\frac{c^2 e^{i\vartheta}}{\pi^2}\,\ds\frac{1}{n^2}\right),
\qquad \lambda\in \partial U_n.\tag 2.24$$
 From (2.24) we get
$$\sqrt{\lambda}\, c=n\pi+\ds\frac{c^2 e^{i\vartheta}}{2\pi}\,\ds\frac{1}{n}+
O\left(\ds\frac{1}{n^3}\right),
\qquad \lambda\in \partial U_n,\quad
n\to+\infty,\tag 2.25$$
where we recall that $\arg(\lambda)\in(-\pi/2,\pi/2].$
Using the trigonometric relations
$$\sin(z+n\pi)=(-1)^n\sin(z),\qquad z\in\bold C,$$
$$\sin(z)=z+o(z),\qquad z\to 0 \text{ in } \bold C,$$
 from (2.25) we get
$$\left|\sin(\sqrt{\lambda}\,c)\right|=\ds\frac{c^2}{2\pi}
\,\ds\frac{1}{n}+o\left(\ds\frac{1}{n}\right),
\qquad \lambda\in \partial U_n,\quad n\to+\infty.\tag 2.26$$
Note that (2.25) implies that
$$n=\ds\frac{\sqrt{\lambda}\,c}{\pi}+O\left(\ds\frac{1}
{\sqrt{\lambda}}\right),
\qquad \lambda\in \partial U_n,\quad \lambda\to\infty,
$$ and hence we can write (2.26) as
$$\left|\sin(\sqrt{\lambda}\,c)\right|=\ds\frac{c}{2|\sqrt{\lambda}|}
+o\left(\ds\frac{1}{\sqrt{\lambda}}\right),
\qquad \lambda\in \partial U_n,\quad
\lambda\to\infty.\tag 2.27$$
The estimate in (2.27), the fact that $\sin(\sqrt{\lambda}\,c)$ grows
exponentially for large $\text{Im}[\sqrt{\lambda}]$ while its
nonzero zeros are confined to the centers
 of the disks $U_n,$ and the minimum modulus principle
applied to the exterior of $U_n$ imply that there exist positive constants
$m$ and $M$ such that
$$\left|\sin(\sqrt{\lambda}\,c)\right|\ge \ds\frac{m}{|\sqrt{\lambda}|},
\qquad \lambda\in\bold C\setminus \ds{\cup_{n=1}^\infty} U_n,\quad |\lambda|\ge M,
\tag 2.28$$
and hence (2.23) and (2.28) yield
$$|h(\lambda)|\le \ds\frac{A\,|\sqrt{\lambda}|}{m}
\,\exp \left(c\,|\lambda|^{(1/2)+\epsilon}\right),\qquad \lambda\in\bold C\setminus
 \ds{\cup_{n=1}^\infty} U_n,\quad |\lambda|\ge M.\tag 2.29$$
On the other hand, by the maximum modulus principle, the maximum of
$|h(\lambda)|$ in the closure of
$U_n$ must occur on the boundary $\partial U_n,$
and hence (2.29) holds, whenever $|\lambda|\ge M,$
perhaps by replacing $A$ there with another positive constant.
Hence, we have proved that the order of $h$ cannot exceed $1/2.$ \qed

\noindent {\bf Proposition 2.7} {\it Let $f$ be an entire
function of $\lambda$ such that}
$$
\cases f(\lambda )=\ds\frac{\exp \left(\left| \text{Im}[\sqrt{\lambda }]\right|
 c\right)}{\sqrt{\lambda }}\,
O(1),\qquad \lambda\to \infty \text { in }\bold C,  \\
\stretch
f\left( \ds\frac{\pi ^{2}n^{2}}{c^{2}}\right) =0,\qquad n\in\bold N,\endcases
\tag 2.30$$
{\it where $c$ is a positive
constant. Then there is a constant $C_{1}$ such that}
$$f(\lambda )=C_{1}\frac{\sin \left(\sqrt{\lambda }\, c\right) }{\sqrt{\lambda }}
=C_{1}\,c\prod_{n=1}^{\infty }\left( 1-\ds\frac{c^{2}\lambda }{\pi ^{2}n^{2}}
\right).\tag 2.31$$
{\it Similarly,
if $g$ is an entire function of $\lambda$ such that}
$$
\cases
g(\lambda )=\exp \left(\left|\text{Im}[\sqrt{\lambda }]\right|c\right)\,O(1),\qquad
\lambda\to \infty \text { in }\bold C,\\
\stretch
g\left( \ds\frac{\pi ^{2}(2n-1)^{2}}{4c^{2}}\right) =0,\qquad
n\in\bold N,\endcases  \tag 2.32$$
{\it then there is there is a constant $C_{2}$ such that}
$$
g(\lambda )=C_{2}\cos\left(\sqrt{\lambda }\, c\right)
=C_{2}\prod_{n=1}^\infty\left( 1-\ds\frac{4c^{2}\lambda }{\pi ^{2}(2n-1)^{2}}
\right).\tag 2.33$$

\noindent PROOF: The second line of (2.30)
implies that $f(\lambda)$ can be written as
$$f(\lambda )=h(\lambda )\,\frac{\sin \left(\sqrt{\lambda }\,c\right) }{\sqrt{
\lambda }},\tag 2.34$$
for some entire function $h(\lambda).$
Using (2.34) in the first
line of (2.30) we get
$$\left|
h(\lambda )\,\frac{\sin \left(\sqrt{\lambda }\,c\right) }{\sqrt{
\lambda }}\right|\le \ds\frac{B\,
\exp \left(\left| \text{Im}[\sqrt{\lambda }]\right|c\right)}
{\left|\sqrt{\lambda }\right|},\qquad \lambda\in\bold C,\tag 2.35$$
for some positive constant $B.$ By Proposition~2.6 we know that
the order of $h$ cannot exceed $1/2.$
On the other hand, by using the exponential representation of
the sine function, as $\lambda\to\infty$ along any ray other than the positive
real axis we have
$$\left|
\frac{\sin \left(\sqrt{\lambda }\,c\right) }{\sqrt{ \lambda
}}\right|=\ds\frac{\exp \left(\left| \text{Im}[\sqrt{\lambda
}]\right|c\right)}{2\left|\sqrt{\lambda
}\right|}\left[1+o(1)\right].\tag 2.36$$ Hence, from (2.35) and
(2.36) we see that $h(\lambda)$ must be bounded on any ray
other than the positive real axis. By invoking a consequence of
the Phragm\'en-Lindel\"of principle (see Theorem 18.1.3 of
[17]) we conclude that $h(\lambda)$ must be a constant, which establishes
(2.31). The proof of (2.33) is obtained in a similar manner.
\qed

\noindent {\bf Proposition 2.8} {\it Let $f$ be an entire
function of $\lambda$ satisfying (2.30), and assume that as
$\lambda\to\infty$ along some fixed ray in the complex plane we
have}
$$
f(\lambda )= \ds\frac{\exp \left(\left| \text{Im}[\sqrt{\lambda
}]\right|c\right)}{\sqrt{\lambda }}\,o(1).\tag 2.37$$ {\it
Then, $f(\lambda)\equiv 0.$ Similarly, let $g$ be an entire
function of $\lambda$ satisfying (2.32), and assume that as
$\lambda\to\infty$ along some fixed ray in the complex plane we
have}
$$
g(\lambda )=\exp \left(\left| \text{Im}[\sqrt{\lambda }]\right|c\right)
\,o(1).\tag 2.38$$
{\it Then, $g(\lambda)\equiv 0.$}

\noindent PROOF: In the proof of Proposition~2.7, 
the further restriction given in (2.37) forces
us to have $C_1=0$ in (2.31), and hence we get
$f(\lambda)\equiv 0.$ Similarly, (2.38) forces to have $C_2=0$
in (2.33), yielding $g(\lambda)\equiv 0.$ \qed

We state a relevant relationship between (2.1) and two
Sturm-Liouville problems in the following corollary.

\noindent {\bf Corollary 2.9} {\it Let $\rho$ satisfy (1.2).
Then, the eigenvalues of the Sturm-Liouville problem}
$$\cases \psi''+\lambda\,\rho(x)\,\psi=0,\qquad 0<x<b,\\
\stretch
\psi(0)=\psi(b)=0,\endcases\tag 2.39$$
{\it exactly correspond
to the zeros of $\phi(b;\lambda),$ where $\phi(x;\lambda)$ is
the unique solution to (2.1). Similarly, the eigenvalues of the
Sturm-Liouville problem}
$$\cases \psi''+\lambda\,\rho(x)\,\psi=0,\qquad 0<x<b,\\
\stretch
\psi(0)=\psi'(b)=0.\endcases\tag 2.40$$ {\it exactly correspond
to the zeros of $\phi'(b;\lambda).$}

The fundamental uniqueness theorem of
inverse spectral theory for Sturm-Liouville problems
indicates that, assuming the existence problem is solved,
the knowledge of two sets of spectra uniquely determines
$\rho.$ It is already known [2,16,19-23] that the combined
knowledge of the eigenvalues of (2.39) and the eigenvalues of
(2.40) uniquely determines $\rho(x)$ for $x\in[0,b].$ Thus, with
the help of Corollary 2.9 we have the following result.

\noindent {\bf Corollary 2.10} {\it Let $\rho_1$ and $\rho_2$
satisfy (1.2), and let $\phi_1(x;\lambda)$ and
$\phi_2(x;\lambda),$ respectively, be the corresponding unique
solutions to (2.1). Then, $\rho_1\equiv \rho_2$ if
$\phi_1(b;\lambda)$ and $\phi_2(b;\lambda)$ have the same set
of zeros and also $\phi'_1(b;\lambda)$ and $\phi'_2(b;\lambda)$
have the same set of zeros.}

\vskip 10 pt
\noindent {\bf 3. THE INVERSE PROBLEM}
\vskip 3 pt

We assume that $\rho$ satisfies (1.2).
The relevant direct problem is the determination of
the special transmission eigenvalues of
(1.1) including
their ``multiplicities" when $\rho(x)$ is known for
$x\in[0,b].$ Conversely, our relevant inverse problem
is the determination of $\rho(x)$ for
$x\in[0,b]$ from the knowledge
of the special transmission eigenvalues of
(1.1) including their ``multiplicities."
 From (2.13) and (2.14) we see that
the direct problem can be equivalently stated as
the determination of the map
$\rho\mapsto \Xi$ and the inverse
problem as the determination of
the map $\Xi\mapsto\rho,$
where $\Xi$ is the quantity appearing in (2.14).
Recall that we are only concerned with the uniqueness aspect of the
inverse problem and not with the existence aspect. In other words, corresponding to our data we assume that
there exists at least one function $\rho$ satisfying (1.2) and we show that
our data leads to a unique $\rho.$

The main conclusion in our paper is that, once the existence
problem is known to be solvable, the function
$\Xi$ uniquely determines $\rho$
in case $a<b,$ where $a$ is the quantity
defined in (1.7).
On the other hand, when $a=b$
it is unclear if $\Xi$ uniquely determines $\rho,$ but we show that
$\Xi$ and $\gamma$ together uniquely determine $\rho,$
where $\gamma$ is the constant appearing in (2.13).
In other words, if $a=b$ then $\rho(x)$ for
$x\in[0,b]$ is uniquely determined by
$D(\lambda)$ for $\lambda\in\bold C.$ First, we present
a special case of the uniqueness result in the following theorem, which also
includes the solution to the relevant existence problem.

\noindent {\bf Theorem 3.1} {\it Assume that $\rho$ satisfies
(1.2), and let the corresponding $D(\lambda)$ be as in
(2.10). If $D(\lambda )\equiv 0$ for $\lambda\in\bold C,$ then
$\rho(x)\equiv 1$ for $x\in[0,b].$}

\noindent PROOF: If $D (\lambda )\equiv 0$ then (2.10)
implies that
$$\ds\frac{\sin (\sqrt{\lambda }\,b)}{\sqrt{\lambda }}\,
\phi'(b;\lambda )=\cos ( \sqrt{\lambda }\,b)\,\phi(b;\lambda
),\qquad \lambda \in \bold C.\tag 3.1$$ Note that each of the
four functions in (3.1), namely, $\sin(\sqrt{\lambda
}\,b)/\sqrt{\lambda},$ $\cos( \sqrt{\lambda }\,b),$
$\phi(b;\lambda),$ and $\phi'(b;\lambda)$ are entire of order
$1/2.$ Furthermore, $\phi(b;\lambda)$ and $\phi'(b;\lambda)$
cannot vanish simultaneously, and $\sin(\sqrt{\lambda
}\,b)/\sqrt{\lambda}$ and $\cos( \sqrt{\lambda }\,b)$ cannot
vanish simultaneously. Thus, (3.1) implies that
$\sin(\sqrt{\lambda }\,b)/\sqrt{\lambda}$ and $\phi(b;\lambda)$ must
have the same set of zeros including multiplicities and
that $\cos( \sqrt{\lambda }\,b)$ and $\phi'(b;\lambda)$ must have
the same set of zeros including multiplicities (note that,
in this particular case, the multiplicities
must all be one). Hence, by the
Hadamard factorization theorem, considering the fact that the
order of each of these four functions is $1/2,$ we must have
$$\phi(b;\lambda)=c_1\,\ds\frac{\sin (\sqrt{\lambda }\,b)}{\sqrt{\lambda }},
\quad
\phi'(b;\lambda)=c_1\,\cos(
\sqrt{\lambda }\,b),\tag 3.2$$
for some nonzero constant $c_1;$ in fact,
(2.16) implies
that $c_1=1.$
By Corollary~2.10 we know that
the combined knowledge of
the zeros of $\phi(b;\lambda)$ and of $\phi'(b;\lambda)$
uniquely determines $\rho.$
Thus, $\rho$ is uniquely determined by
the combined knowledge of
the zeros of $\sin (\sqrt{\lambda}\,b)/\sqrt{\lambda}$
and of $\cos(\sqrt{\lambda}\,b),$ and it is already known that
those combined zeros correspond to
$\rho(x)\equiv 1$ for $x\in[0,b].$ \qed

In the next theorem, we present our uniqueness result when $a<b.$

\noindent {\bf Theorem 3.2} {\it Assume that
for the function $\Xi$ appearing in (2.14) there
corresponds at least one function $\rho$ satisfying (1.2);
assume also that
$a<b,$ where $a$ is the quantity defined in (1.7).
Then, $\rho$ is uniquely determined by
$\Xi;$ in other words, the knowledge of special transmission
eigenvalues of (1.1) with ``multiplicities" uniquely determines
$\rho.$}

\noindent PROOF: Let us assume that
$\rho_1$ and $\rho_2$ correspond to $\Xi_1$ and $\Xi_2,$
respectively, and let $D_1(\lambda)$ and
$D_2(\lambda)$ be the corresponding quantities in (2.13) with
$\gamma_1$ and $\gamma_2$ being the respective constants there.
We will show that $\rho_1\equiv\rho_2$ if
$\Xi_1\equiv \Xi_2.$ Let us also use $\phi_1$ and $\phi_2$ to denote the
solutions to (2.1) corresponding to $\rho_1$ and $\rho_2,$ respectively.
 From the line above (1.8) we see that
if $\Xi_1(\lambda)\equiv \Xi_2(\lambda)$ then
$a_1=a_2,$ where
$a_1$ and $a_2$ are the corresponding
quantities for $\rho_1$ and $\rho_2,$ respectively.
Let us use $a$ to denote the common value of
$a_1$ and $a_2.$ Since we assume that $a<b,$ by (2.5),
(2.6), (2.8), and (2.9) we have
$$\phi_1(b;\lambda)=
\ds\frac{\exp \left(\left| \text{Im}[\sqrt{\lambda }]\right|
a\right)}{\sqrt{\lambda }}\,
O(1),\quad
\phi_2(b;\lambda)=
\ds\frac{\exp \left(\left| \text{Im}[\sqrt{\lambda }]\right|
a\right)}{\sqrt{\lambda }}\,
O(1),\qquad \lambda\to \infty \text { in }\bold C,\tag 3.3$$
$$\phi_1(b;\lambda)=
\ds\frac{\exp \left(\left| \text{Im}[\sqrt{\lambda }]\right|
b\right)}{\sqrt{\lambda }}\,
o(1),\quad
\phi_2(b;\lambda)=
\ds\frac{\exp \left(\left| \text{Im}[\sqrt{\lambda }]\right|
b\right)}{\sqrt{\lambda }}\,
o(1)
,\qquad \lambda\to \infty \text { in }\bold C_{\varepsilon},\tag 3.4$$
$$\phi'_1(b;\lambda)=
\exp \left(\left| \text{Im}[\sqrt{\lambda}]\right|
a\right)\,
O(1),\quad
\phi'_2(b;\lambda)=
\exp \left(\left| \text{Im}[\sqrt{\lambda}]\right|
a\right)\,
O(1),
\qquad \lambda\to \infty \text { in }\bold C,\tag 3.5$$
$$\phi'_1(b;\lambda)=
\exp \left(\left| \text{Im}[\sqrt{\lambda}]\right|
b\right)\,
o(1),\quad
\phi'_2(b;\lambda)=
\exp \left(\left| \text{Im}[\sqrt{\lambda}]\right|
b\right)\,
o(1),
\qquad \lambda\to \infty \text { in }\bold C_{\varepsilon},\tag 3.6$$
where
$\bold C_{\varepsilon}$ is the sector defined in (2.7). Since
$\Xi_1(\lambda)\equiv\Xi_2(\lambda),$ from (2.13) and (2.14) it follows that
$$\ds\frac{1}{\gamma_1}\,D_1\left(\ds\frac{n^2\pi^2}{b^2}\right)=\ds\frac{1}{\gamma_2}\,
D_2\left(\ds\frac{n^2\pi^2}{b^2}\right),
\qquad n\in\bold N,
\tag 3.7$$
and hence from (2.10) we get
$$\ds\frac{1}{\gamma_1}\,\phi_1\left(b;\ds\frac{n^2\pi^2}{b^2}\right)=
\ds\frac{1}{\gamma_2}\,\phi_2\left(b;\ds\frac{n^2\pi^2}{b^2}\right),
\qquad n\in\bold N.
\tag 3.8$$
In a similar way, with the help of (2.9), (2.10), (2.13), and (2.14), by using
$$\Xi_1\left(b;\ds\frac{(2n-1)^2\pi^2}{4b^2}\right)=
\Xi_2\left(b;\ds\frac{(2n-1)^2\pi^2}{4b^2}\right),
\qquad n\in\bold N,
\tag 3.9$$
we obtain
$$\ds\frac{1}{\gamma_1}\,\phi'_1\left(b;\ds\frac{(2n-1)^2\pi^2}{4b^2}\right)
=\ds\frac{1}
{\gamma_2}\,\phi_2'\left(b;\ds\frac{(2n-1)^2\pi^2}{4b^2}\right),
\qquad n\in\bold N.
\tag 3.10$$
In Proposition~2.8 by choosing $c=b$ and
$f(\lambda)=\phi_1(b;\lambda)/\gamma_1-\phi_2(b;\lambda)/\gamma_2,$
we see (3.3), (3.4), and (3.8) imply that
$f(\lambda)\equiv 0.$ Similarly, in Proposition~2.8 by choosing
$c=b$ and
$g(\lambda)=\phi'_1(b;\lambda)/\gamma_1-\phi_2'(b;\lambda)/\gamma_2,$
we see that (3.5), (3.6), and (3.10) imply that
$g(\lambda)\equiv 0.$ On the other hand,
$f(\lambda)\equiv 0$ indicates that
$\phi_1(b;\lambda)$ and $\phi_2(b;\lambda)$ have the same set of zeros,
and $g(\lambda)\equiv 0$ indicates that
$\phi'_1(b;\lambda)$ and $\phi'_2(b;\lambda)$ have the same set of zeros.
Thus, using Corollary~2.10 we conclude that $\rho_1\equiv\rho_2.$ \qed

The next uniqueness theorem applies to the case $a=b.$

\noindent {\bf Theorem 3.3} {\it Assume that
for the function $\Xi$ appearing in (2.14) there
corresponds at least one function $\rho$ satisfying (1.2);
assume also that
$a=b,$ where $a$ is the quantity defined in (1.7).
Then, $\rho$ is uniquely determined by the combined knowledge of
$\Xi$ and the constant $\gamma$ appearing in
(2.13); in other words, the knowledge of special transmission
eigenvalues of (1.1) with ``multiplicities" along with the knowledge of
$\gamma$ uniquely determines
$\rho.$}

\noindent PROOF: The proof is similar to the proof of
Theorem~3.2 with appropriate modifications we indicate here.
As in the proof of Theorem~3.2 we have
$\Xi_1(\lambda)\equiv \Xi_2(\lambda),$ but we also
have $\gamma_1=\gamma_2,$ and we want to show that
$\rho_1\equiv\rho_2.$ By Corollary~2.10 it is sufficient to prove that
$\phi_1(b;\lambda)$ and $\phi_2(b;\lambda)$ have the same set of zeros
and that
$\phi'_1(b;\lambda)$ and $\phi'_2(b;\lambda)$ have the same set of zeros.
Since $a=b,$ this time we have
(3.3) and (3.5), but not (3.4) or (3.6). Proceeding as in (3.7)-(3.10) verbatim,
and in Proposition~2.8 by choosing $f$ and $g$ as in the proof of Theorem~3.2, we obtain
$$\ds\frac{1}{\gamma_1}\,
\phi_1(b;\lambda)-\ds\frac{1}{\gamma_2}\,\phi_2(b;\lambda)
=C_1\,\ds\frac{\sin(\sqrt{\lambda}\,b)}{\sqrt{\lambda}},\tag 3.11$$
$$\ds\frac{1}{\gamma_1}\,
\phi'_1(b;\lambda)-\ds\frac{1}{\gamma_2}\,\phi_2'(b;\lambda)
=C_2\,\cos(\sqrt{\lambda}\,b),\tag 3.12$$
for some constants $C_1$ and $C_2.$
Evaluating (3.11) and (3.12) at $\lambda=0$ and
using (2.16), we get
$$C_1=C_2=\ds\frac{1}{\gamma_1}-\ds\frac{1}{\gamma_2}.\tag 3.13$$
Since we assume $\gamma_1=\gamma_2,$ we see from (3.13) that $C_1=C_2=0.$
Thus, from (3.11) and (3.12) we get
$\phi_1(b;\lambda)=\phi_2(b;\lambda)$ and
$\phi'_1(b;\lambda)=\phi'_2(b;\lambda),$
indicating that
$\phi_1(b;\lambda)$ and $\phi_2(b;\lambda)$ have the same set of zeros
and also that
$\phi'_1(b;\lambda)$ and $\phi'_2(b;\lambda)$ have the same set of zeros. \qed

Having considered the inverse problem when $a<b$ and $a=b$
in Theorems~3.2 and 3.3, respectively,
let us now comment on the case $a>b.$ The method we use
to prove the uniqueness for $a\le b$ does not apply to the case $a>b,$ as the following
argument clarifies.
The lack of applicability of our technique
to the case $a>b$
certainly does not mean that a uniqueness result does not exist
when $a>b.$ The unique recovery of
$\rho$ from $D(\lambda)$ defined in (2.10)
is based on our ability to extract each of
$\phi(b;\lambda)$ and $\phi'(b;\lambda)$ up to a constant multiplicative
factor. When $a>b$ let us the consider the determination of
two functions
$\phi(b;\lambda)$ and $\phi'(b;\lambda)$
that are entire in $\lambda$ and of order $1/2$ and that satisfy
the respective asymptotics related to (2.5) and (2.6),
namely, as $\lambda\to \infty \text { in }\bold C$
$$
\phi(b;\lambda)=\ds\frac{\exp \left(\left| \text{Im}[\sqrt{\lambda }]\right|
a\right)}{\sqrt{\lambda }}\,
O(1),
\quad
\phi'(b;\lambda)=\exp \left(\left| \text{Im}[\sqrt{\lambda }]\right|
a\right)\,
O(1),
\tag 3.14$$
for which
$$\ds\frac{\sin (\sqrt{\lambda }\,b)}{\sqrt{\lambda }}
\,\phi'(b;\lambda)-\cos (\sqrt{
\lambda }\,b)\,\phi(b;\lambda)=D (\lambda ).\tag 3.15$$
Let $\zeta$ be any entire function of $\lambda$ having the asymptotics
$$\zeta(\lambda)=\exp \left(\left| \text{Im}[\sqrt{\lambda }]\right|
(a-b)\right)\,
O(1),
\qquad \lambda\to \infty \text { in }\bold C.\tag 3.16$$
Letting
$$\check\phi(b;\lambda):=\phi(b;\lambda)+
\ds\frac{\sin (\sqrt{\lambda }\,b)}{\sqrt{\lambda }}
\,\zeta(\lambda),
\quad
\check\phi'(b;\lambda):=\phi'(b;\lambda)+
\cos (\sqrt{\lambda }\,b)
\,\zeta(\lambda),$$
we see that $\check\phi(b;\lambda)$ and
$\check\phi'(b;\lambda)$ are entire in $\lambda$
and that (3.14) and (3.15) are satisfied
when we replace in them $\phi(b;\lambda)$ with
$\check\phi(b;\lambda)$ and replace $\phi'(b;\lambda)$ with
$\check\phi'(b;\lambda).$ Because of (3.16), $\zeta(\lambda)$
must be a constant when $a=b$ and must be zero when $a<b,$ but
no such restrictions exist when $a>b.$ Thus, our method does
not allow us to conclude the unique determination of $\rho$
from $D(\lambda)$ when $a>b.$

\vskip 10 pt
\noindent {\bf 4. THE INVERSE PROBLEM FOR THE SCHR\"ODINGER EQUATION}
\vskip 3 pt

In the case of the Schr\"{o}dinger operator, the interior transmission
eigenvalue problem is analogous to the corresponding problem for
the wave equation with variable speed.
Instead of (1.1), we have
$$
\cases
-\Delta \tilde\Psi + V(\bold x) \,\tilde\Psi=\mu \tilde\Psi,\qquad \bold x\in \Omega,\\
\stretch
-\Delta \tilde\Psi_0 = \mu \tilde\Psi_0,\qquad \bold x\in \Omega, \\
\stretch
\tilde\Psi=\tilde\Psi_0,\quad
\displaystyle\frac{\partial \tilde\Psi}{\partial \bold n}
=\frac{\partial \tilde\Psi_0}{\partial \bold n},
\qquad \bold x\in \partial \Omega, \endcases \tag 4.1$$
where $\mu$ is the spectral parameter,
$V(\bold x)$ is a real-valued potential that is square integrable
on $\Omega,$ and it is assumed that
$V(\bold x)\equiv 0$ outside $\Omega.$ Those $\mu$-values yielding nontrivial
solutions $\tilde\Psi$ and $\tilde\Psi_0$ to (4.1)
are called {\it transmission eigenvalues} of (4.1).
In the spherically-symmetric case, using $V(x)$ instead of
$V(\bold x)$ with $x:=|\bold x|,$ we have the following analog of
Proposition~1.1. We omit its proof because it is similar
to the proof of Proposition~1.1.

\noindent {\bf Proposition 4.1} {\it Consider the
special case of (4.1) with
$\Omega$ being the three-dimensional ball of
radius $b$ centered at the origin,
where only
spherically-symmetric wave functions are allowed
and it is assumed that such wave functions
are continuous in the closure of $\Omega.$ Then, the
corresponding transmission eigenvalues of
(4.1) coincide with the eigenvalues of the boundary-value problem}
$$\cases -\tilde\Phi''+V(x) \,\tilde\Phi=\mu\, \tilde\Phi
,\qquad 0<x<b,\\
\stretch
\tilde\Phi(0) = 0, \quad
\displaystyle\frac{\sin (\sqrt{\mu}\,b)}{\sqrt{\mu}}\,
\tilde\Phi'(b)-\cos (\sqrt{\mu}\,b)\,\tilde\Phi(b)=0.
\endcases\tag 4.2$$

The eigenvalues of (4.2), namely the $\mu$-values for
which (4.2) has a nontrivial solution
coincide with the {\it special transmission eigenvalues}
of (4.1), namely
those transmission eigenvalues
of (4.1) for which the
corresponding wavefunctions are spherically symmetric
in addition to $V$ being spherically symmetric.
Note that the boundary condition
at $x=b$ in (4.2) suggests an analog of $D(\lambda)$ appearing in (2.10).
We define
$$\tilde D (\mu) := \frac{\sin (\sqrt{\mu}\,b)}{\sqrt{\mu}}
\,\tilde\phi'(b;\mu)-\cos (\sqrt{\mu}\,b)\,
\tilde\phi(b;\mu),\tag 4.3$$
where $\tilde\phi(x;\mu)$ is the analog
of $\phi(x;\lambda)$ appearing
in (2.4) and is
the unique solution to
the initial-value problem
$$\cases -\tilde \phi''+V(x)\,
\tilde \phi=\mu\, \tilde \phi,
\qquad 0<x<b,\\
\stretch
\tilde \phi(0)=0,\quad
\tilde \phi'(0)=1.\endcases\tag 4.4$$

The following proposition contains results that
are analogous to those stated in
Propositions~2.1, 2.2, and 2.3.

\noindent {\bf Proposition 4.2} {\it Assume that
$V$ is real valued and square integrable
on $[0,b].$ Then, for each
$\mu\in\bold C,$ (4.4) has
a unique solution $\tilde\phi(x;\mu).$
For each fixed $\mu,$ the functions
$\tilde\phi(\cdot;\mu)$ and $\tilde\phi'(\cdot;\mu)$
cannot vanish at the same $x$-value.
For each fixed $x\in(0,b],$
the quantities $\tilde\phi(x;\cdot)$ and
$\tilde\phi'(x;\cdot)$
are entire in $\mu$ and there exists a positive constant $A$ such that}
$$
\left|\tilde\phi(x;\mu)-\frac{1}{\sqrt{
\mu}}\,\sin \left(\sqrt{\mu}\,x
\right)\right|\leq \frac{A}{\left|\sqrt{\mu}\right|}\,
\exp \left( \left|\text{Im}[\sqrt{\mu}]\right|\,x \right),$$
$$\left|\tilde\phi'(x;\mu)-
\cos \left( \sqrt{\mu}\,x \right)
\right| \leq A\exp \left( \left|\text{Im} [\sqrt{\mu}]\right|
\,x\right).$$
{\it Furthermore,
for each fixed $x\in [0,b],$ as $\mu\to\infty$
in the sector $\bold C_{\varepsilon}$ defined in (2.7),
we have}
$$
\tilde\phi(x;\mu)=\frac{1}{\sqrt{\mu}}
\,\sin \left( \sqrt{\mu}\,x\right)\left[
1+O\left(\frac{1}{\sqrt{\mu }}\right) \right],$$
$$
\tilde\phi'(x;\mu )=\cos
\left( \sqrt{\mu}\,x \right) \left[
1+O\left( \frac{1}{\sqrt{\mu}}\right) \right].$$

We note that (4.2), (4.3), and (4.4) are closely related.
If $\mu_j$ is an eigenvalue of (4.2)
with an eigenfunction $\tilde \Phi(x;\mu_j),$ then
$\tilde \Phi(x;\mu_j)$ must be a constant
multiple of $\tilde\phi(x;\mu_j),$ where
$\tilde\phi(x;\mu)$ denotes the unique solution to (4.4).
Hence, from (4.3) we conclude that
$\tilde D(\mu_j)=0.$ Thus, with the help of
Proposition~4.1 we conclude that the special transmission
eigenvalues of (4.1), the eigenvalues of (4.2), and the zeros of
(4.4) all coincide. On the other hand, there exists only one
linearly independent eigenfunction for a given eigenvalue $\mu_j$ of
(4.2) whereas the multiplicity of $\mu_j$ as a zero of
$\tilde D(\mu)$ may be greater than one. We will refer to
the multiplicity of a zero $\mu_j$ of $\tilde D(\mu)$ as the
``multiplicity" of the special transmission eigenvalue $\mu_j$
and also as the ``multiplicity" of the eigenvalue $\mu_j$ of
(4.2).

Note that from (4.3) and the second line
of (4.4) we obtain
$$\tilde D(0)=b\,\tilde\phi'(b;0)-\tilde\phi(b;0),$$
and, contrary to (2.11), generically
we have $\tilde D(0)\ne 0,$ although we may
have $\tilde D(0)=0$ for some potentials. For example,
if $V(x)\equiv 0,$ then we have
$\tilde\phi(x;\mu)=\sin(\sqrt{\mu}\,x)/\sqrt{\mu},$
yielding $\tilde D(0)=0.$ In fact,
$\tilde D(\mu)\equiv 0$ for 
$\mu\in\bold C$ in that special case.

Our goal in this section is to show that $V(x)$ for $0<x<b$
is uniquely determined by the corresponding
$\tilde D(\mu)$ known for all $\mu\in\bold C.$
In fact, we will see that, up to the multiplicative constant
$\tilde\gamma$ appearing in (4.5),
$\tilde D(\mu)$ is uniquely determined by
the knowledge of its zeros
including the multiplicities of those zeros. Since those zeros are exactly
the eigenvalues of (4.2), we will conclude that
the knowledge of
the eigenvalues of (4.2) including their
``multiplicities" and the value of $\tilde\gamma$
uniquely determines $V.$ Since the eigenvalues
of (4.2) are the special transmission eigenvalues of
(4.1), we will also
conclude that the knowledge
of those special transmission eigenvalues
including their ``multiplicities"
and $\tilde\gamma$ uniquely determines
$V.$ Since the proofs are similar to those in the case of
the variable-speed wave equation studied
in the previous sections, we will omit some of
the proofs.

As in the case of the variable-speed wave equation, we only consider
the uniqueness aspect of our inverse problem and not the existence
aspect. In other words, corresponding to our data $\tilde D(\mu)$
or its equivalents, we assume that there exists at least one
potential $V,$ where $V(x)$ is real valued and belongs to
$L^2(0,b).$ We then prove that if $V_1$ and $V_2$ are two
such potentials, then we must have $V_1\equiv V_2.$
Let us also clarify that the equality $V_1\equiv V_2$ is meant
to be an equality in the almost-everywhere sense
because we deal with potentials in the class
$L^2(0,b)$ whereas the corresponding
equality $\rho_1\equiv\rho_2$ obtained
in Section~3 holds pointwise because $\rho_1$ and $\rho_2$
satisfy (1.2).

The following theorem summarizes the properties of $\tilde D(\mu)$
defined in (4.3), and it is an analog of Theorem~2.4.
We omit the proof because it is similar to
the proof of Theorem~2.4. 

\noindent {\bf Proposition 4.3} {\it Assume that
$V$ is real valued and square integrable
on $[0,b].$ Then, the quantity
$\tilde D(\mu)$ defined in (4.3) is entire
in $\mu$ and its order does not exceed 1/2.
Thus, by the Hadamard factorization theorem
$\tilde D(\mu)$ is determined,
uniquely up to a multiplicative constant, from its zeros
as}
$$\tilde D(\mu)=\tilde\gamma\,
\mu^{\tilde d}\prod_{n=1}^{\infty }\left( 1-\frac{
\mu}{\mu _{n}}\right),\tag 4.5$$
{\it with $\mu_n$ for $n\in\bold N$
being the nonzero
zeros of
$\tilde D(\mu),$ some of which may be repeated, and
$\tilde d$ denoting the multiplicity
of the zero as a zero of $\tilde D(\mu).$}

The results stated in the
following theorem are analogous to those
stated in Corollaries~2.9 and 2.10.

\noindent {\bf Theorem 4.4} {\it Assume that
$V$ is real valued and square integrable
on $[0,b],$ and let $\tilde \phi(x;\mu)$ denote the unique solution
to (4.4). We then have the following:}
\item{(i)} {\it The zeros of $\tilde \phi(b;\mu)$ coincide with the eigenvalues of
the Sturm-Liouville problem}
$$\cases -\tilde \psi''+V(x)\,\tilde\psi=\mu\,\tilde \psi,\qquad 0<x<b,\\
\stretch
\tilde\psi(0)=\tilde\psi(b)=0.\endcases\tag 4.6$$
\item{(ii)} {\it The zeros of $\tilde \phi'(b;\mu)$ coincide with the eigenvalues of
the Sturm-Liouville problem}
$$\cases -\tilde \psi''+V(x)\,\tilde\psi=\mu\,\tilde \psi,\qquad 0<x<b,\\
\stretch
\tilde\psi(0)=\tilde\psi'(b)=0.\endcases\tag 4.7$$
\item{(iii)} {\it The data consisting of the
eigenvalues of (4.6) and (4.7)
uniquely determines $V$ if the existence is assured. In other words, assuming that
there exists at least one $V$ corresponding to the data, if $V_1$ and $V_2$ correspond
to the same data then we must have $V_1(x)\equiv V_2(x)$ on $[0,b].$}
\item{(iv)} {\it The data consisting of the
zeros of $\tilde\phi(b;\mu)$ and $\tilde\phi'(b;\mu)$
uniquely determine $V$ if the existence is assured. In other words, assuming that
there exists at least one $V$ corresponding to the data, if $V_1$ and $V_2$ correspond
to the same data then we must have $V_1(x)\equiv V_2(x)$ on $[0,b].$}

\noindent PROOF: We obtain (i) and (ii) by
comparing (4.4) and (4.6) and by noting that $\tilde\phi(0)=\tilde\psi(0)=0.$
We note that (iii) is a version of the well-known uniqueness
result by Borg [3]. Finally, (iv) is a consequence of (i)-(iii). \qed

The following is the analog of the uniqueness result stated in Theorem~3.3.

\noindent {\bf Theorem 4.5} {\it Assume that $V(x)$ is real valued
and square integrable on $[0,b].$ Then, $V$ is uniquely determined by the function
$\tilde D(\mu)$ appearing in (4.3) if we assume that
there exists at least one $V$ corresponding to $\tilde D.$
Equivalently stated, if the existence is assured,
$V$ is uniquely determined by the knowledge
of the special transmission eigenvalues of (4.1) with their ``multiplicities"
and the
constant $\tilde\gamma$ appearing in (4.5).}

\noindent PROOF: If $V_1$ and $V_2$ correspond to
$\tilde D_1(\mu)$ and $\tilde D_2(\mu),$ then we need to show that
$V_1\equiv V_2$ when $\tilde D_1(\mu)\equiv \tilde D_2(\mu).$
Let $\tilde\phi_1(x;\mu)$ and $\tilde\phi_2(x;\mu)$ be the solutions to
(4.4) corresponding to $V_1$ and $V_2,$ respectively.
Because of Theorem~4.4 (iv), it is sufficient
to show that $\tilde\phi_1(b;\mu)=\tilde\phi_2(b;\mu)$
and $\tilde\phi'_1(b;\mu)=\tilde\phi'_2(b;\mu),$ which is proved
by proceeding as in the proof of Theorem~3.3. \qed

By Proposition~4.3 we know that the knowledge of
$\tilde D(\mu)$ is equivalent to the knowledge of
its zeros with their multiplicities and the
constant $\tilde\gamma$ appearing in (4.5).
We have already seen that the zeros of
$\tilde D(\mu),$ the eigenvalues of (4.2), and
the special transmission eigenvalues of (4.1)
all coincide. Thus, from Theorem~4.5 we obtain
the following corollary.

\noindent {\bf Corollary 4.6} {\it Assume that $V(x)$ is real valued
and square integrable on $[0,b].$ Assuming that there exists at least
one $V$ corresponding to the data, $V$ is uniquely determined
by the data consisting of the zeros of $\tilde D(\mu)$ in (4.5)
with their multiplicities and
the constant $\tilde\gamma$ there.
Equivalently, assuming the existence, $V$ is uniquely determined
by the data consisting of the eigenvalues of (4.2)
with their ``multiplicities" and
the constant $\tilde\gamma.$}

One consequence of Corollary~4.6 is that
if $\tilde D(\mu)\equiv 0,$ then $V(x)\equiv 0,$
which is the analog of Theorem~3.1.

Let us mention that it is an open problem whether the value of
$\tilde\gamma$ appearing in (4.5) can be determined from the
zeros of $\tilde D(\mu).$ If the answer is yes, then $\tilde\gamma$
is not needed for the unique
determination of $V,$ and as seen from Corollary~4.6
the zeros of $\tilde D(\mu)$ with their multiplicities would be
sufficient for that purpose. The technique we use to prove
the uniqueness assumes the knowledge of $\tilde\gamma,$
but this does not rule out the possibility that
there might be another method to obtain the
uniqueness from the data consisting only of
the zeros of $\tilde D(\mu)$ and their
multiplicities. We note that
in the discrete version of the inverse transmission
problem for the Schr\"odinger equation,
$\tilde\gamma$ is determined [29] in the generic
case by the zeros of
$\tilde D(\mu)$ and their multiplicities.

\vskip 10 pt

\noindent {\bf Acknowledgments.} The authors thank Prof. Paul Sacks of Iowa State University
for his comments and suggestions. The first author has been partially supported by
the Texas NHARP under grant
no. 003656-0046-2007 and by
DOD-BC063989; he is grateful for the hospitality he received
during a recent visit to the National Technical University of Athens.
The second and third authors have been partially supported by a $\Pi$.E.B.E. grant from the National Technical University of Athens.

\vskip 10 pt

\noindent {\bf{References}}

\vskip 3 pt

\item{[1]}
L. V. Ahlfors,
{\it Complex analysis,} 3rd ed., McGraw-Hill, New York, 1979.

\item{[2]}
V. Barcilon, {\it Explicit solution of the inverse problem for a
vibrating string,}  J. Math. Anal. Appl. {\bf 93}, 224--234 (1983).

\item{[3]}
G. Borg, {\it Eine Umkehrung der Sturm-Liouvilleschen
Eigenwertaufgabe,}  Acta Math. {\bf 78}, 1--96 (1946).

\item{[4]}
F. Cakoni, M. {\c C}ay\"oren, and D. Colton,
{\it Transmission eigenvalues and the nondestructive testing of dielectrics,}
Inverse Problems {\bf 24}, 065016 (2008).

\item{[5]}
F. Cakoni, D. Colton, and D. Gintides,
{\it The interior transmission problem,}
in: A. Charalambopoulos, D. I. Fotiadis, and D. Polyzos (eds.),
{\it Advanced topics in scattering theory and biomedical engineering: Proceedings of the 9th international workshop on mathematical methods in scattering theory and biomedical engineering,} World Scientific Publ., Singapore, 2010, pp. 368--380.

\item{[6]}
F. Cakoni, D. Colton, and H. Haddar,
{\it
The computation of lower bounds for the norm of the index of refraction in an anisotropic media from far field data,} J. Integral Equations Appl. {\bf 21}, 203--227 (2009).

\item{[7]}
F. Cakoni, D. Colton, and H. Haddar,
{\it
The interior transmission problem for regions with cavities,}
SIAM J. Math. Anal. {\bf 42},
145--162 (2010).

\item{[8]}
F. Cakoni, D. Colton, and H. Haddar,
{\it On the determination of Dirichlet or transmission eigenvalues from far field data,}
C. R. Math. Acad. Sci. Paris {\bf 348}, 379--383 (2010).

\item{[9]}
F. Cakoni, D. Colton, and P. Monk,
{\it
On the use of transmission eigenvalues to estimate the index of refraction from far field data,} Inverse Problems {\bf 23}, 507--522 (2007).

\item{[10]}
F. Cakoni and D. Gintides,
{\it
New results on transmission eigenvalues,} Inverse Probl. Imaging {\bf 4}, 39--48 (2010).

\item{[11]}
F. Cakoni, D. Gintides, and H. Haddar,
{\it The existence of an infinite discrete set of transmission eigenvalues,} SIAM J. Math. Anal. {\bf 42}, 237--255 (2010).

\item{[12]}
F. Cakoni and H. Haddar,
{\it
On the existence of transmission eigenvalues in an inhomogeneous medium,}
Appl. Anal. {\bf 88}, 475--493 (2009).

\item{[13]}
D. Colton and R. Kress,
{\it Inverse acoustic and electromagnetic scattering theory,}
2nd ed.,  Springer,
New York, 1998.

\item{[14]}
D. Colton and P. Monk,
{\it The inverse scattering problem for time-harmonic acoustic waves in an inhomogeneous medium,}
Quart. J. Mech. Appl. Math. {\bf 41}, 97--125 (1988).

\item{[15]}
D. Colton, L. P\"aiv\"arinta, and J. Sylvester,
{\it The interior transmission problem,}
Inverse Probl. Imaging {\bf 1}, 13--28 (2007).

\item{[16]}
H. Dym and H. P. McKean, {\it Gaussian processes, function theory, and the inverse spectral problem,}
Dover Publ., New York, 2008.

\item{[17]}
E. Hille, {\it Analytic function theory,}
Volume II, AMS Chelsea Publishing, Providence, R.I., 2005.

\item{[18]}
A. Kirsch, {\it On the existence of transmission eigenvalues,}
Inverse Probl. Imaging {\bf 3}, 155--172 (2009).

\item{[19]}
M. G. Krein,
{\it Determination of the density of a nonhomogeneous symmetric cord by its frequency spectrum,}
Doklady Akad. Nauk SSSR (N.S.) {\bf 76}, 345--348 (1951) [Russian].

\item{[20]}
M. G. Krein,
{\it On inverse problems for a nonhomogeneous cord,} Doklady Akad. Nauk SSSR (N.S.)
{\bf 82}, 669--672 (1952) [Russian].

\item{[21]}
M. G. Krein,
{\it On some new problems of the theory of oscillations of Sturmian systems,}
Akad. Nauk SSSR. Prikl. Mat. Meh. {\bf 16}, 555--568 (1952) [Russian].

\item{[22]}
N. Levinson,
{\it The inverse Sturm-Liouville problem,}
Mat. Tidsskr. B {\bf 25}, 25--30 (1949).

\item{[23]}
B. M. Levitan and M. G. Gasymov,
{\it Determination of a
differential equation by two of its spectra,} Russian Math.
Surveys {\bf 19}, 1--63 (1964).

\item{[24]}
J. R. McLaughlin and P. L. Polyakov,
{\it On the
uniqueness of a spherically symmetric speed of sound from transmission
eigenvalues,} J. Differential Equations {\bf 107}, 351--382 (1994).

\item{[25]}
J. R. McLaughlin, P. L. Polyakov, and P. E. Sacks,
{\it Reconstruction of a spherically symmetric speed of sound,}
SIAM J. Appl. Math. {\bf 54}, 1203--1223 (1994).

\item{[26]}
J. R. McLaughlin, P. E. Sacks, and M. Somasundaram,
{\it Inverse scattering in acoustic media using interior transmission eigenvalues,}
in: G. Chavent, G. Papanicolaou, P. Sacks, and W. Symes (eds.),
{\it Inverse problems in wave propagation,} Springer,
New York, 1997, pp. 357--374.

\item{[27]}
M. A. Naimark, {\it Linear differential operators,}
Parts I and II, Frederick Ungar Publishing, New York, 1967 and 1968.

\item{[28]}
L. P\"aiv\"arinta and J. Sylvester,
{\it Transmission eigenvalues,} SIAM J. Math. Anal.
{\bf 40}, 738--753 (2008).

\item{[29]} V. G. Papanicolaou and A. V. Doumas,
{\it On the discrete one-dimensional inverse transmission eigenvalue problem,}
Inverse Problems {\bf 27}, 015004 (2011).

\item{[30]}  J. P\"{o}schel and E. Trubowitz, {\it Inverse
spectral theory,} Academic Press, Boston, 1987.

\item{[31]}
B. P. Rynne and B. D. Sleeman, {\it The interior transmission problem and inverse scattering from inhomogeneous media,} SIAM J. Math. Anal. {\bf 22}, 1755--1762 (1991).

\end